\documentclass[prd, %twocolumns,
showpacs,nofootinbib,preprintnumbers]{revtex4}
\usepackage{amsmath}
\usepackage{amssymb}
\usepackage{amsfonts}
\usepackage{graphicx}
\usepackage{dcolumn}
\usepackage{hyperref}
\newcommand{\be}{\begin{equation}}
\newcommand{\ee}{\end{equation}}
\newcommand{\bea}{\begin{eqnarray}}
\newcommand{\eea}{\end{eqnarray}}
\newcommand{\beaa}{\begin{eqnarray*}}
\newcommand{\eeaa}{\end{eqnarray*}}

\newcommand{\nn}{\nonumber \\}
\newcommand{\e}{{\rm e}}

\begin{document}

\title{Cardy-Verlinde formula in FRW Universe with inhomogeneous
generalized fluid and dynamical entropy bounds near the future
singularity}

\author{Iver Brevik$^1$, Shin'ichi Nojiri$^2$, Sergei D.
Odintsov$^{3,4}$\footnote{Also at Tomsk State
Pedagogical University}, Diego S\'aez-G\'omez$^4$}

\affiliation{$^1$Department of Energy and Process Engineering,
Norwegian University of Science and Technology, N-7491 Trondheim,
Norway}
\affiliation{$^2$Department of Physics, Nagoya University, Nagoya
464-8602, Japan}
\affiliation{$^3$Instituci\`{o} Catalana de Recerca i Estudis
Avan\c{c}ats (ICREA), Barcelona}
\affiliation{$^4$Institut de Ciencies de l'Espai (IEEC-CSIC),
Campus UAB, Facultat de Ciencies, Torre C5-Par-2a pl, E-08193
Bellaterra (Barcelona), Spain}

\begin{abstract}

We derive a formula for the entropy  for a multicomponent coupled
fluid, which under special conditions reduces to the
Cardy-Verlinde form relating the entropy of a closed FRW universe
to its energy together with  its  Casimir energy. The generalized
fluid obeys an inhomogeneous equation of state. A viscous  dark
fluid is included, and  also
 modified gravity is included in terms of its  fluid representation. It
is demonstrated how  such an expression reduces to the standard
Cardy-Verlinde formula corresponding to the 2d CFT
entropy  in some special cases. The dynamical entropy bound for a
closed FRW universe with  dark components is obtained.
The universality of the  dynamical entropy bound near a
 future singularity (of all known four types), as well as near the Big Bang
singularity, is investigated. It is demonstrated that except from
some special cases  of Type II and Type IV singularities the
dynamical entropy bound is violated near  the singularity even if
 quantum effects are taken into account. The dynamical entropy bound seems to be
universal for the case of a regular universe, including the asymptotic de Sitter universe.
\end{abstract}

\maketitle

\section{Introduction}

Recent astronomical data indicate that the current universe is expanding
with cosmic acceleration caused by the so-called Dark Energy (DE).
The well-known $\Lambda$CDM model where Dark Energy comes from the
effective cosmological constant fits quite well the observational
data. Nevertheless, it is not excluded that the current DE may have a
phantom origin (or  will become phantom-like DE in the near future).
Moreover, the possibility of a quintessence-like DE with  effective
equation of state (EoS) parameter being very close to $-1$ is widely
discussed in the recent literature.
It is quite likely that the current FRW universe is spatially-flat.
However, the occurrence of a slightly spatially-curved FRW universe is
 not excluded by observational bounds.

The study of phantom-like or quintessence-like effective dark fluids
opens for a number of new phenomena which are typical for such a DE
universe.
For instance, it is known that phantom DE may drive the future
universe to a so-called finite-time Big Rip singularity (for earlier
works on this, see Refs.\cite{ref5,Elizalde:2004mq}). From another
side, quintessence-like DE may  bring the future universe to a
milder future singularity (like the sudden singularity
\cite{barrow,singularity} where the effective energy-density
is finite).
Actually, the study of Ref.\cite{ConformalAnomaly1} shows that there
are four different types of future finite-time singularities where the
Type I singularity corresponds to the Big Rip, the sudden singularity is of
Type II, etc. The universe looks quite strange near to the
singularity where curvature may grow up so that
quantum gravity effects may be dominant \cite{Elizalde:2004mq}.
In any case, the study of the universe under critical conditions (for
instance, near a future singularity) may clarify the number of
fundamental issues relating seemingly different physical
theories.

Some time ago \cite{Verlinde} it was shown that the first FRW equation for a
closed FRW universe may have a more fundamental origin than what is
expected from standard General Relativity. It was demonstrated that
this equation may be rewritten so as to describe the universe entropy in terms
of total energy and Casimir energy (the so-called Cardy-Verlinde (CV)
formula). Moreover, it turns out that the corresponding formula has a
striking correspondence with the Cardy formula for the entropy
of a two-dimensional conformal field theory (2d CFT).
Finally, the formula may be rewritten as a dynamical entropy bound
from which a number of  entropy bounds, proposed earlier,
follow. The connection between the  standard gravitational equation and the 2d CFT
 dynamical entropy bound indicates a very deep relation
between gravity and thermodynamics. It raises  the question about
to which extent  the CV formula is universal. Problems of this
sort are natural to study when the universe is under critical
conditions, such as near a singularity.

The present work is devoted to a study of the universality of the
CV formula and the corresponding dynamical entropy bound in a DE
universe filled with a generalized fluid, especially near the
singularity regime. Generalization of the CV formula for a
multicomponent fluid with interactions, assuming the  EoS to be
inhomogeneous, is presented. The viscous case is incorporated. It
is shown that the standard CV formula with correct power (square
root) is restored only for some very special cases. The dynamical
entropy bound for such fluids near  all the four types of the
future singularity is considered. It is demonstrated that this
dynamical entropy bound is most likely violated near the
singularity (except from some cases of Type II and Type IV
singularity). This situation is not qualitatively changed even if
account is taken  of quantum effects in conformally invariant
theory. Using the formalism of modified gravity to describe an
effective dark fluid, the corresponding CV formula is constructed
also for $F(R)$-gravity. The corresponding dynamical entropy bound
is derived. It is shown that the bound is satisfied for a  de
Sitter universe solution. Further discussion and outlook  is given
in the discussion section.

\section{Generalization of Cardy-Verlinde formula in FRW Universe for various types of fluids}

This section is devoted to consideration of the Cardy-Verlinde
formula for more general scenarios than those considered in
previous works (see \cite{Verlinde}-\cite{Youm}). We consider a
$(n+1)$-dimensional spacetime described by the FRW metric, written
in comoving coordinates as
\be ds^2=-dt^2+\frac{a(t)^2
dr^2}{1-kr^2}+r^2 d\Omega^2_{n-1}\ , \label{1} \ee
where $k=-1, 0, +1$  for an open, flat, or closed spatial
Universe, $a(t)$ is taken to have unit of length, and $d\Omega^2_{n-1}$ is the metric of an $n-1$ sphere. By inserting the
metric (\ref{1}) in the Einstein field equations the FRW equations
are derived,
\be H^2 = \frac{16\pi
G}{n(n-1)}\sum^m_{i=1}\rho_i-\frac{k}{a^2}\, ,\quad \dot{H} =
-\frac{8\pi G}{n-1}\sum^m_{i=1}(\rho_i + p_i) + \frac{k}{a^2}\ .
\label{3} \ee
Here $\rho_i={E_i}/{V}$ and $p_i$ are the
energy-density and pressure of the matter component $i$ that fills
the Universe. In this paper we consider only the $k=1$ closed
Universe. Moreover, we assume an equation of state (EoS) of the
form $p_i=w_i\rho_i$ with $w_i$ constant for each fluid, and
assume at first no interaction between the different components.
Then, the conservation law for energy has the form \be
\dot{\rho_i}+nH \left(\rho_i+p_i\right)=0\ , \label{4} \ee and by
solving (\ref{4}), we find that the $i$ fluid depends on the scale
factor as \be \rho_i\propto a^{-n(1+w_i)}\ . \label{5} \ee Let us
now review the case of Ref.\cite{Youm}, where just one fluid with
EoS $p=w\rho$ and $w= \text{constant}$ is considered. The total
energy inside the comoving volume $V$, $E=\rho V$, can be written
as the sum of an extensive part $E_\mathrm{E}$ and a subextensive
part $E_\mathrm{C}$, called the Casimir energy, and  takes the
form: \be E(S,V)=E_\mathrm{E}(S,V)+\frac{1}{2}E_\mathrm{C}(S,V)\ .
\label{6} \ee Under a rescaling of the entropy
($S\rightarrow\lambda S$) and the volume ($V\rightarrow\lambda
V$), the extensive and subextensive parts of the total energy
transform as \be E_\mathrm{E}(\lambda S,\lambda V) = \lambda
E_\mathrm{E}(S,V)\ , \quad E_\mathrm{C}(\lambda S,\lambda V) =
\lambda^{1-2/n} E_\mathrm{C}(S,V)\ . \label{7} \ee Hence, by
assuming that the Universe satisfies the first law of
thermodynamics, the term corresponding to the Casimir energy
$E_\mathrm{C}$ can be seen as a violation of the Euler identity
according to the definition in Ref.\cite{Verlinde}: \be
E_\mathrm{C}=n(E+pV-TS) \ . \label{8} \ee Since the total energy
behaves as $E\sim a^{-nw}$ and by the definition (\ref{6}), the
Casimir energy also goes as $E_\mathrm{C}\sim a^{-nw}$. The FRW
Universe expands adiabatically, $dS=0$, so the products
$E_\mathrm{C}a^{nw}$ and $E_\mathrm{E}a^{nw}$ should be
independent of the volume $V$, and be just a function of the
entropy. Then, by the rescaling properties (\ref{7}), the
extensive and subextensive part of the total energy can be written
as functions of the entropy only \cite{Youm}, \be E_\mathrm{E}=
\frac{\alpha}{4\pi a^{nw}}S^{w+1}\ , \quad
E_\mathrm{C}=\frac{\beta}{2\pi a^{nw}}S^{w+1-2/n}\ . \label{9} \ee
Here $\alpha$ and $\beta$ are undetermined constants. By combining
these expressions with (\ref{6}), the entropy of the Universe is
written as a function of the total energy $E$ and the Casimir
energy $E_\mathrm{C}$, \cite{Youm}, \be S=\left(\frac{2\pi
a^{nw}}{\sqrt{\alpha\beta}}\sqrt{E_\mathrm{C}(2E-E_\mathrm{C})}
\right)^{\frac{n}{n(w+1)-1}}\ , \label{10} \ee which for $w=1/n$
(radiation-like fluid) reduces to \cite{Verlinde}, \be
S=\frac{2\pi
a}{\sqrt{\alpha\beta}}\sqrt{E_\mathrm{C}(2E-E_\mathrm{C})}\ ,
\label{19} \ee which has the same form as the Cardy formula given
in Ref.\cite{Cardy}. The first FRW equation (\ref{3}) can be
rewritten as a relation between thermodynamics variables, and
yields \be
S_\mathrm{H}=\frac{2\pi}{n}a\sqrt{E_\mathrm{BH}(2E-E_\mathrm{BH})}\
, \quad \mbox{where} \quad S_\mathrm{H}=(n-1)\frac{HV}{4G}\, ,
\quad E_\mathrm{BH}=n(n-1)\frac{V}{8\pi Ga^2}\ . \label{20} \ee It
is easy to check that for the bound proposed in
Ref.\cite{Verlinde}, $E_\mathrm{C}\leq E_\mathrm{BH}$, the
equation for the entropy (\ref{19}) coincides with the first FRW
equation (\ref{20}) when the bound is reached. We will see below
that when there are several fluid components, the same kind of
expression as in Ref.\cite{Verlinde} cannot be found. Nor is there
the same correspondence with the FRW equation when the bound is
saturated.

\subsection{Multicomponent Universe}

If $m$ fluids are considered with arbitrary EoS, $p_i=w_i\rho_i$,
the expression for the total entropy is simple to derive just by
following the same method as above. The total entropy is given by
the sum of the entropies for each fluid, \be S=\sum_{i=1}^m S_i =
\sum_{i=1}^m \left(\frac{2\pi
a^{nw_i}}{\sqrt{\alpha\beta}}\sqrt{E_{iC}(2E_{i}-E_{iC})}
\right)^{\frac{n}{n(w_i+1)-1}}\ . \label{11} \ee This expression
cannot be reduced to one depending only on the total energy unless
very special conditions on the nature of the fluids are assumed.
Let us   for simplicity  assume that  there are
only two fluids with EoS given by $p_1=w_1\rho_1$ and
$p_2=w_2\rho_2$, $w_1$ and $w_2$ being constants. We can
substitute the fluids by an effective fluid described by the EoS
\be p_\mathrm{eff}=w_\mathrm{eff}\rho_\mathrm{eff}\ , \quad
\text{where} \quad w_\mathrm{eff}=
\frac{p_1+p_2}{\rho_1+\rho_2}=w_1+\frac{w_2-w_1}{1+\rho_1/\rho_2}\
, \label{12} \ee and $p_\mathrm{eff}=\frac{1}{2}(p_1+p_2)$,
$\rho_\mathrm{eff}=\frac{1}{2}(\rho_1+\rho_2)$. Then, by using the
energy conservation equation (\ref{4}), we find $\rho_1\sim
(a/a_0)^{-n(1+w_1)}$ and $\rho_2\sim (a/a_0)^{-n(1+w_2)}$, where $a_0$ is assumed to be the value of the scale factor at the time $t_0$. The effective
EoS parameter $w_\mathrm{eff}$ can be expressed as a function of
the scale factor $a(t)$ \be
w_\mathrm{eff}=w_1+\frac{w_2-w_1}{1+(a/a_0)^{n(w_2-w_1)}}\ . \label{13}
\ee The total energy inside a volume $V$ becomes \be E_T=E_1+E_2
\propto (a/a_0)^{-nw_1}+(a/a_0)^{-nw_2}\ . \label{16} \ee As the
energy is proportional to two different powers of the scale factor
$a$, it is not possible to write it as a function of
the total entropy only. As a special case, if the EoS parameters are $w_1=w_2=w_\mathrm{eff}$, the formula for the entropy
reduces to (\ref{10}), and coincides with the CV formula  when $w_\mathrm{eff}=1/n$. \\

As another  case, one might consider that for some epoch of the cosmic history,  $w_1 \gg w_2$. Taking also   $a>>a_0$, we could then approximate  the total energy  by the
function $E_T \propto a^{-nw_2}$. From  (\ref{6}) the
Casimir energy would also depend on the same power of $a$,
$E_\mathrm{C}\propto a^{-nw_2}$. The expression (\ref{10}) is again recovered with $w=w_2$. \\

Thus in general, when a multicomponent FRW Universe is assumed, the formula for the total entropy does
not resemble the Cardy formula, nor does it correspond to the FRW
equation when the Casimir bound is reached.  It becomes possible to
reconstruct the formula (\ref{19}), and establish the correspondence with the
Cardy formula, only if we make specific choices for the EoS of the fluids.

\subsection{Interacting fluids}

As a second case we now consider a Universe, described by the metric (\ref{1}),
 filled with two interacting fluids. One can write the
energy conservation equation for each fluid as
\be
\dot{\rho_1}+n H(\rho_1+p_1)= Q\ , \quad
\dot{\rho_2}+n H(\rho_2+p_2)= -Q\ ,
\label{22}
\ee
where $Q$ is a function that accounts for the
energy exchange between the fluids. This kind of interaction has
been discussed previously in studies of dark energy and dark
matter. The effective EoS parameter is given by the same expression
(\ref{12}) as before. With a
specific choice for the coupling function $Q$, the equations
(\ref{22}) may be solved. One can in principle find the
dependence of the energy densities $\rho_{1,2}$ on the scale
factor $a$,
\be
\rho_1= a(t)^{-n(1+w_1)}\left(C_1+\int
a^{n(1+w_1)}Q(t)dt \right) \ , \quad
\rho_2= a(t)^{-n(1+w_2)}\left(C_2-\int a^{n(1+w_2)}Q(t)dt \right)\ ,
\label{24}
\ee
where $C_1$ and $C_2$ are integration constants.
In general it is not possible to reproduce the CV
formula, and the result will be a sum of  different
contributions, similar to the entropy expression given in (\ref{11}).
However, for the case where the effective EoS parameter (\ref{13})
is a constant, the expression for the entropy will be given by
 equation (\ref{10}) as before. This condition only holds when $w_\mathrm{eff}=w_1=w_2$, where the situation is thus  equivalent to the one-fluid case, and the entropy reduces to the CV formula when $w_\mathrm{eff}=1/n$.   \\

Let us  consider a simple choice for the function $Q$ that leads to the CV formula for a certain limit. Let
 $Q=Q_0 a^mH$, where $m$ is a positive number,
$Q_0$ is a constant, and $H(t)$  the Hubble parameter.
Then the integral in (\ref{24}) is easily calculated,
and the energy densities depend on the scale factor according to
\be
\rho_1=C_1a^{-n(1+w_1)}+k_1a^m\ , \quad
\rho_2=C_2a^{-n(1+w_2)}+k_2a^m\ ,
\label{24c}
\ee
where
$k_{1,2}=Q_0/(n(1+w_{1,2})+m)$. If we restrict ourselves to the
regime where $a\gg C_{1,2}$ such that the first terms in the expressions
for $\rho_{1,2}$ are negligible, the effective EoS parameter becomes
\be
w_\mathrm{eff}=w_1+\frac{w_2-w_1}{1 + \frac{k_1}{k_2}}\ .
\label{24d}
\ee
Then, the entropy of the universe is given by (\ref{10}) with
$w=w_\mathrm{eff}$. The CV formula can be reproduced only with very  specific
choice of the free parameters, just as above.

We have thus shown that in general  a formula for the entropy of the type (\ref{10}) cannot be reconstructed for interacting fluids. Coincidence with the Cardy formula is obtained  if the effective EoS parameter is  radiation-like,
$w_\mathrm{eff}=1/n$. Then the expression for the entropy turns out to be
in agreement with the formula (\ref{19}), corresponding to
the first FRW equation (\ref{20})
when the Casimir energy reaches the bound
$E_\mathrm{C}=E_\mathrm{BH}$.

\subsection{Inhomogeneous EoS fluid and bulk viscosity}

Let us now explore the case of an $n+1$-dimensional Universe filled with
a fluid satisfying an inhomogeneous EoS. This kind of EoS,  generalizing  the perfect fluid model, has been
considered in several papers as a way to describe effectively the
dark energy (see \cite{InhEoS} and \cite{Interac.Fluids2}).
We assume an EoS expressed as a function of the scale factor,
\be
p=w(a)\rho+g(a)\ .
\label{28}
\ee
This EoS fluid could be taken to correspond to  modified gravity,
or to bulk viscosity (Ref.\cite{InhEoS}). By introducing (\ref{28})
in the energy conservation equation (\ref{4}) we obtain
\be
\rho'(a)+\frac{n(1+w(a))}{a}\rho(a)=-n\frac{g(a)}{a}\ .
\label{29}
\ee
Here we have performed a variable change $t=t(a)$ such that
the prime over $\rho$ denotes derivative with respect to the
scale factor $a$. The general solution of this equation is
\be
\rho(a)=\e^{-F(a)}\left(K -n\int\e^{F(a)}\frac{g(a)}{a}da\right)
\quad \text{where} \quad F(a)=\int^a \frac{1+w(a')}{a'}da'\ ,
\label{30}
\ee
and $K$ is an integration constant. As shown above,
only for some special choices of the functions $w(a)$ and $g(a)$,
the formula (\ref{19}) can be recovered. Let us assume, as  an example,  that
$w(a)=-1$ and $g(a)=-a^m$, with $m= $ constant.
Then, the energy density behaves as $\rho\propto a^m$. Hence, by following the same steps as described above, the
extensive and subextensive energy go as $a^{m+n}$, and by imposing
conformal invariance and the rescaling properties (\ref{7}), we
calculate the dependence on the entropy to be
\be
E_\mathrm{E}=
\frac{\alpha}{4\pi na^{-(m+n)}}S^{-m/n}\, , \quad
E_\mathrm{C}=\frac{\beta}{4\pi na^{-(m+n)}}S^{-(2n+m/n)}\ .
\label{31}
\ee
The expression for the entropy is easily
constructed by combining these two expressions and substituting
the extensive part by the total energy. This gives us the same expression as in (\ref{10}) with $w=-(n+m)/n$. Note that for
$m=-(1+n)$, the formula (\ref{19}) is recovered and also its
correspondence with the CFT formula. However for a generic power
$m$, the CV formula cannot be reconstructed, like
the cases studied above. Only for some special choices does the correspondence
work, leading to the identification between the FRW equation and
the Cardy formula. \\

Let us now consider an inhomogeneous EoS fluid due to bulk
viscosity. From a hydrodynamical perspective it is natural to
extend the formalism so as to incorporate viscosity effects.
Working to the first order in the deviations from thermal
equilibrium we are faced with two viscosity coefficients, namely
the shear viscosity $\eta$ and the bulk viscosity $\zeta$. In
conformity with spatial isotropy we shall assume, as usual, that
only the bulk viscosity contributes. (For a review of viscous
cosmology and entropy, one may consult Ref.\cite{brevik05} and also
Refs.\cite{SDO-IB1}.) The
viscous fluid may be considered as a special kind of inhomogeneous
EoS fluid, although of a different kind from that of
Eq.~(\ref{28}) above. We set the number of spatial dimensions $n$
equal
to $3$. The energy-momentum tensor can be written as
\begin{equation}
T_{\mu\nu}=\rho U_\mu U_\nu+\tilde{p}h_{\mu\nu}\ ,
\end{equation}
where $h_{\mu\nu}=g_{\mu\nu}+U_\mu U_\nu$ is the projection tensor
and
\begin{equation}
\tilde{p}=p-3H\zeta\, ,
\end{equation}
the effective pressure. For simplicity let us assume in this
subsection the following simple fluid model:
\begin{equation}
w={\rm constant}, \quad g=0, \quad \zeta= {\rm constant,}
\end{equation}
In comoving coordinates, $U^0=1,U^i=0$.
The first of the FRW equations (\ref{3}) maintains its form (it is
viscous insensitive), whereas the second equation becomes
\begin{equation}
\dot{H}=-4\pi G(\rho+\tilde{p})+\frac{1}{a^2}\ ,
\end{equation}
showing that $\tilde{p}$ is now the thermodynamically important
pressure. We see that the relationship $p=w\rho$, or
\begin{equation}
\tilde{p}=w\rho -3H\zeta\ ,
\end{equation}
can be considered as an EoS in the present case.

Introduction of a viscosity means effectively the introduction of a
length parameter, and so the conformal invariance of the formalism is
lost.
The question arises: Can the entropy arguments leading to the
Cardy-Verlinde formula be carried over to the viscous case? The
answer actually turns out to be affirmative, at least when $\zeta$
is small. The most delicate point in the line of arguments is
the assumed pure entropy dependence of the product $Ea^{3w}$.
Now, if one uses the FRW equations to derive the
``energy equation'' ($k=+1$ assumed)
\begin{equation}
\frac{d}{dt}\left(\rho a^{3(1+w)}\right)=9\zeta H^2a^{3(1+w)}\ ,
\end{equation}
one can combine this with the equation for entropy production
\begin{equation}
N\dot{\sigma}=\frac{9\zeta}{T}H^2\ ,
\end{equation}
where $N$ is the particle (baryon) density and $\sigma$ the
entropy per particle. As discussed in some detail in
\cite{brevik05} it follows that, since the total energy $E\sim
\rho a^3$ and the total entropy $S\sim N\sigma a^3$, the quantity
$Ea^{3w}$ becomes independent of the volume $V$ and is a function
of $S$ only. This generalizes the pure entropy dependence of the
product $Ea$ found by Verlinde in the case of a non-viscous
radiation dominated universe. We obtain the expression (\ref{10}) with $n=3$
as the generalized Cardy-Verlinde formula and reducing to the
standard formula
(with square root) when the universe is radiation dominated.

The following point ought finally to be noted. The energy
conservation equation ${T^{0\nu}}_{;\nu}=0$ implies
\begin{equation}
\dot{\rho}+3H(\rho+{\tilde p})=0\ ,
\end{equation}
so that equations (21) can be taken over to the viscous case, only
with the substitutions $p_i\rightarrow \tilde{p}_i$.

\section{On the cosmological bounds near future singularities}

In Ref.\cite{Verlinde}, Verlinde proposed a new universal bound on
cosmology based on a restriction of the Casimir energy $E_C$; cf.
 his entropy formula (\ref{19}). This new bound postulated was
\be
E_\mathrm{C}\leq E_\mathrm{BH}\ ,
\label{B1}
\ee
where $E_\mathrm{BH}=n(n-1)\frac{V}{8\pi Ga^2}$. It was
deduced by the fact that in the limit when the Universe passes
between strongly and weakly self-gravitating regimes, the
Bekenstein entropy $S_\mathrm{B}=\frac{2\pi a}{n}E$ and the
Bekenstein-Hawking entropy $S_\mathrm{BH}=(n-1)\frac{V}{4Ga}$,
which define each regime, are equal. This bound could be
interpreted to mean that the Casimir energy  never  becomes  able
to reach sufficient energy, $E_\mathrm{BH}$, to form a black hole
of the size of the Universe. It is easy to verify that the strong
($Ha\geq 1$) and weak ($Ha\leq 1$) self-gravity regimes have the
following restrictions on the total energy,
\bea
E &\leq &
E_\mathrm{BH}\ \quad \text{for} \quad Ha\leq 1 \ , \nn
E &\geq &
E_\mathrm{BH}\ \quad \text{for} \quad Ha\geq 1\ .
\label{B2}
\eea
From here it is easy to calculate the bounds on the entropy of
the Universe in the case when the Verlinde formula (\ref{19}) is
valid; this is (as  shown in the above sections) for an
effective radiation dominated Universe $w_\mathrm{eff}\sim 1/n$.
The bounds for the entropy deduced in Ref.\cite{Verlinde} for
$k=1$ are
\bea
S &\leq& S_\mathrm{B}\ \quad \text{for} \quad Ha\leq 1 \ , \nn
S &\geq& S_\mathrm{H}\ \quad \text{for} \quad Ha\geq 1\ ,
\label{B3}
\eea
where $S_\mathrm{B}$ is the Bekenstein entropy defined
above, and $S_\mathrm{H}$ is the Hubble entropy given by
(\ref{20}).
Note that for the strong self-gravity regime, $Ha\geq 1$,
the energy range is $E_\mathrm{C}\leq E_\mathrm{BH}\leq E$.
According to the formula (\ref{19}) the maximum entropy is
reached when the bound is saturated, $E_\mathrm{C}=E_\mathrm{BH}$.
Then $S=S_\mathrm{H}$, such that the FRW equation coincides
with
the CV formula, thus indicating  a connection with
CFT.  For the weak regime, $Ha\leq 1$, the range of energies
goes as $E_\mathrm{C}\leq E\leq E_\mathrm{BH}$ and the maximum
entropy is reached earlier, when $E_\mathrm{C}=E$, yielding the
result $S=S_\mathrm{B}$. The entropy bounds can be extended to more
general cases, corresponding to an arbitrary EoS parameter $w$. By taking
the bound (\ref{B1}) to be universally valid one can easily
deduce the new entropy bounds for each regime, from the expression
of the entropy (\ref{10}). These new bounds, discussed in
Ref.\cite{Youm}, differ from the ones given in (\ref{B3}), but
still establish a bound on the entropy as long as the bound on
$E_\mathrm{C}$ expressed in (\ref{B1}) is taken to be valid. The
entropy bounds can be related through  the first FRW equation,
yielding the following quadratic expression (for $k=1$),
\be
S^2_\mathrm{H}+(S_\mathrm{B}-S_\mathrm{BH})^2=S^2_\mathrm{B}\ .
\label{B3a}
\ee
We would like to study
what happens to the bounds,  particularly to the fundamental
bound (\ref{B1}), when the cosmic evolution is close to a future
singularity; then the effective fluid  dominating the
cosmic evolution could have an unusual EoS. As  shown below,
for some class of future singularities such a bound could soften
the singularities in order to avoid  violation of the
\textit{universal} bound (\ref{B1}). It  could be interpreted
to mean that  quantum effects become important when the bound is reached.
 However, as the
violation of the bound could happen long before the singularity
even in the presence of  quantum effects, it could be a signal
of  breaking of the universality of the bound (\ref{B1}). Let
us first of all give a list of the possible future cosmic
singularities, which can be classified according to
Ref.\cite{ConformalAnomaly1} as
\begin{itemize}
\item Type I (``Big Rip''): For $t\rightarrow t_s$, $a\rightarrow
\infty$ and $\rho\rightarrow \infty$, $|p|\rightarrow \infty$.
\item Type II (``Sudden''): For $t\rightarrow t_s$, $a\rightarrow
a_s$ and $\rho\rightarrow \rho_s$, $|p|\rightarrow \infty$.
(see Refs.\cite{singularity,barrow})
\item Type III: For $t\rightarrow t_s$, $a\rightarrow a_s$
and $\rho\rightarrow \infty$, $|p|\rightarrow \infty$.
\item Type IV: For $t\rightarrow t_s$, $a\rightarrow a_s$ and
$\rho\rightarrow \rho_s$, $p \rightarrow p_s$
but higher derivatives of Hubble parameter diverge.
\end{itemize}
Note that the above list was suggested in the case of a flat FRW
Universe.
As we consider in this paper a closed Universe ($k=1$), we should
make an analysis to see
if the list of singularities given above is also valid in this case.
It is straightforward to see that all the singularities listed above
can be reproduced for a particular choice of the effective EoS.
To show how the cosmic bounds behave for each type of singularity, we
could write an explicit
solution of the FRW equations, expressed as a function of time
depending on free parameters
that will be fixed for each kind of singularity. Then, the Hubble
parameter may be written as follows
\be
H(t)=\sqrt{\frac{16\pi G}{n(n-1)}\rho -\frac{1}{a^2}}=H_1(t_s-t)^m
+H_0\ ,
\label{B4}
\ee
where $m$ is a constant properly chosen for each type of singularity.
Note that this is just a solution that ends in the singularities
mentioned above, but there are other solutions which also reproduce
such singularities.  We will study how the cosmic bounds behave near
each singularity listed above.

As pointed out in
Ref.\cite{Elizalde:2004mq,ConformalAnomaly1}, around a singularity
 quantum effects could become important as the curvature of the
Universe grows and diverges in some of the cases. In other words, approaching the finite-time future singularity the curvature grows and universe reminds the early universe where quantum gravity effects are dominant ones because of extreme conditions.
Then one has to take into account the role of such quantum gravity effects which should define the behaviour of the universe just before the singularity. Moreover, they may act so that to prevent the singularity occurrence. In a sense, one sees the return of quantum gravity era.
However, the consistent quantum gravity theory does not exists so far.
Then, in order to estimate the influence of quantum effects to universe near to singularity one can use the effective action formulation.
We will apply the effective action produced by conformal anomaly (equivalently, the effective fluid with pressure/energy-density corresponding to conformal anomaly ones) because of several reasons.
It is known that at high energy region (large curvature) the conformal invariance is restored so one can neglect the masses. Moreover, one can use large N approximation to justify why large number of quantum fields may be considered as effective quantum gravity. Finally, in the account of quantum effects via conformal anomaly we keep explicitly the graviton (spin 2) contribution. The conformal anomaly $T_A$ has the
following well-known form
\be
T_A=b\left(F+\frac{2}{3}\square R\right) +b'G+b''\square R\ .
\label{B4a}
\ee
Here we assume for simplicity a 3+1 dimensional spacetime. Then, $F$
is the square of a 4D Weyl tensor and $G$ is the Gauss-Bonnet
invariant,
\be
F=\frac{1}{3}R^2-2R_{ij}R^{ij}+R_{ijkl}R^{ijkl}\ ,\quad
G=R^2-4R_{ij}R^{ij}+R_{ijkl}R^{ijkl}\, .
\label{B4b}
\ee
The coefficients $b$ and $b'$ in (\ref{B4a}) are described by the
number of $N$ scalars, $N_{1/2}$ spinors, $N_{1}$ vector fields,
$N_2$ gravitons and $N_\mathrm{HD}$ higher derivative conformal
scalars.
They can be written as
\be
b=\frac{N+6N_{1/2}+12N_{1}+611N_{2}-8N_\mathrm{HD}}{120(4\pi)^2}\
,\quad
b'=-\frac{N+11N_{1/2}+62N_{1}+1411N_{2}-28N_\mathrm{HD}}{360(4\pi)^2}\
.
\label{B4c}
\ee
As $b''$ is arbitrary  it can be shifted by a finite
renormalization of the local counterterm. The conformal anomaly $T_A$
can be written as $T_A=-\rho_A+3p_A$, where $\rho_A$ and $p_A$ are
the energy and pressure densities respectively. By using (\ref{B4a})
and the energy conservation equation $\rho_A+3H(\rho_A+p_A)=0$, one
obtains the following expression for $\rho_A$
\cite{ConformalAnomaly1,ConformalAnomaly2},
\bea
\rho_A &=& -\frac{1}{a^4}\int dt a^4 HT_A \nn
&=& -\frac{1}{a^4}\int dt
a^4H\left[-12b\dot{H}^2+24b'(-\dot{H}^2+H^2\dot{H}+H^4)
�-(4b+6b'')(\dddot{H}+7H\ddot{H}+4\dot{H}^2+12H^2\dot{H}) \right]\ .
\label{B4d}
\eea
The quantum corrected FRW equation is given by
\be
H^2=\frac{8\pi G}{3}(\rho+\rho_A)-\frac{1}{a^2}\ .
\label{B4e}
\ee
We study now how the bounds behave around the singularity in
the classical case when no quantum effects are added, and then
include the conformal anomaly (\ref{B4a}) quantum effects in the
FRW equations. We will see that for some cases the violation of the
cosmic bound can be avoided.

\subsection{Big Rip Singularity}

This type of singularity has been very well studied and has become
very popular
as it is a direct consequence in the majority of the cases when the
effective EoS parameter is less than $-1$, the so-called phantom
case \cite{ref5,Elizalde:2004mq}.
Observations currently indicate that the phantom barrier could have
already been crossed
or it will be crossed in the near future, so a lot of attention has
been paid to this case.
It can be characterized by the solution (\ref{B4}) with $m\leq -1$,
and this yields
 the following dependence of the total energy density on the scale
factor near the singularity,
when $a\gg 1$, for a closed Universe ($k=1$),
\be
\rho=\frac{n(n-1)}{16\pi G}H^2+\frac{1}{a^2}\sim a^{-n(1+w)} \quad
\text{for} \quad t\rightarrow t_s\ ,
\label{B5}
\ee
where we have chosen $H_1=2/n|1+w|$ with $w<-1$, $m=-1$ and $H_0=0$
for clarity.
This solution drives the Universe to a Big Rip singularity for
$t\rightarrow t_s$,
where the scale factor diverges. If the singularity takes place, the
bound (\ref{B1})
has to be violated before this happens. This can be seen from
equation (\ref{B5}),
as the Casimir energy behaves as $E_\mathrm{C}\propto a^{n|w|}$ while
the Bekenstein-Hawking energy
goes as $E_\mathrm{BH}\propto a^{n-2}$.
Then, as $w<-1$, the Casimir energy grows faster than the BH energy,
so close to the singularity
where the scale factor becomes very big, the value of $E_\mathrm{C}$ will be
much higher than $E_\mathrm{B}$, thus violating
the bound (\ref{B1}).
Following the postulate from Ref.\cite{Verlinde} one could interpret
the bound (\ref{B1})
as the limit where General Relativity and Quantum Field Theory
converge,
such that when the bound is saturated quantum gravity effects should
become important.
QG corrections could help to avoid the violation of the bound and may
be the Big Rip singularity occurrence.
As this is just a postulate based on the CV formula, which is only
valid for special cases
as  shown in the sections above, the bound on $E_\mathrm{C}$
could not be valid for any kind of fluid.

Let us now include the
conformal anomaly (\ref{B4a}) as a quantum effect that becomes
important around the Big Rip. In such a case there is a phase
transition and the Hubble evolution will be given by the solution of
the FRW equation (\ref{B4e}). Let us  approximate to get some
qualitative results,  assuming $3+1$ dimensions.
Around $t_s$ the curvature is large, and
$|\rho_A|>>(3/\kappa^2)H^2+k/a^2$. Then  $\rho\sim-\rho_A$, and from
(\ref{B4d})we get
\be
\dot{\rho}+4H\rho=H\left[-12b\dot{H}^2+24b'(-\dot{H}^2+H^2\dot{H}+H^4)
�-(4b+6b'')(\dddot{H}+7H\ddot{H}+4\dot{H}^2+12H^2\dot{H}) \right]\ .
\label{B5a}
\ee
We assume that the energy density, which diverges in the
classical case, behaves now as
\be
\rho\sim (t_s-t)^{\lambda}\ ,
\label{B5b}
\ee
where $\lambda$ is some negative number. By using the energy
conservation equation $\dot{\rho}+3H(1+w)\rho=0$, the Hubble
parameter goes as $H\sim 1/(t_s-t)$. We can check if this
assumption is correct in the presence of  quantum effects by
inserting both results in Eq. (\ref{B5a}). We get
\be
\rho\sim 3H^4(-13b+24b')\ .
\label{B5c}
\ee
Hence as $b>0$ and $b'<0$,  $\rho$ becomes negative,
which is an unphysical result. Thus  $\rho$ should not
go to infinity in the presence of the quantum correction. This
is the same result as obtained in Ref.\cite{ConformalAnomaly1} where
numerical analysis showed that the singularity is moderated by the
conformal anomaly, so that the violation of the bound that naturally
occurs in the classical case can be avoided/postponed when quantum
effects are included.

\subsection{Sudden singularity}

This kind of singularity is also problematic with respect to the bounds, but as
the energy density $\rho$ does not diverge,
the violation of the bound may be avoided for some special choices.
The sudden singularity can be described
by the solution (\ref{B4}) with $0<m<1$, and constants $H_{0,1}>0$.
Then the scale factor goes as
\be
a(t)\propto \exp\left[ -\frac{H_1}{m+1}(t_s-t)^{m+1}+H_0t\right] \ ,
\label{B6}
\ee
which gives $a(t)\sim \e^{h_0t}$ (de Sitter) close to $t_s$. From the
first FRW equation the total energy density becomes
\be
\rho=H^2(t)+\frac{1}{a^2}=\left[H_1(t_s-t)^m+H_0 \right]^2
+ \exp\left[ 2\frac{H_1}{m+1}(t_s-t)^{m+1}-2H_0t\right]\ ,
\label{B6a}
\ee
which tends to a constant $\rho\sim H_0^2+\e^{-2H_0t_s}$ for
$t\rightarrow t_s$.
Then the Casimir energy grows as $E_\mathrm{C}\propto H^2_0
a^{n}+a^{n-2}$, while $E_\mathrm{BH}\propto a^{n-2}$ close to $t_s$.
The BH energy grows slower than the Casimir energy, and the bound is
violated for a finite $t$.
However, by an specific choice of the coefficients, the violation of
the bound (\ref{B1}) could be avoided.
For $H_0=0$, and by some specific coefficients, the bound could be
obeyed.
In general, it is very possible that $E_\mathrm{C}$ exceeds
its bound. In the presence of quantum corrections, the singularity
can be avoided but the bound can still be violated, depending on the
free parameters for each model. We may assume that in the presence of
the conformal anomaly for $n=3$, the energy density grows as
\cite{ConformalAnomaly1}
\be
\rho=\rho_0+\rho_1(t_s-t)^{\lambda}\ ,
\label{B6b}
\ee
where $\rho_0$ and $\rho_1$ are constants, and $\lambda$ is now a
positive number. Then the divergences on the higher derivatives of
the Hubble parameter can be avoided, as  is shown in
Ref.\cite{ConformalAnomaly1}. Nevertheless, $E_\mathrm{C}$ still
grows faster than $E_\mathrm{BH}$, such that the Universe has to be
smaller than a critical size in order to hold the bound (\ref{B1}) as
 is pointed in Ref.\cite{Youm} for the case of a vacuum dominated
universe.

\subsection{Type III singularity}

This type of singularity is very similar to the Big Rip, in spite of
the scale factor $a(t)$ being finite at the singularity.
The solution (\ref{B4}) reproduces this singularity by taking
$-1<m<0$. The scale factor goes as
\be
a(t)= a_s\exp \left[-\frac{H_1}{m+1}(t_s-t)^{m+1} \right]\ ,
\label{B7}
\ee
where for simplicity we take $H_0=0$. Then, for $t\rightarrow t_s$,
the scale factor $a(t)\rightarrow a_s$.
To see how $E_\mathrm{C}$ behaves near the singularity, let us write
it in terms of the time instead of the scale factor,
\be
E_\mathrm{C}\propto a_s^nH_1^2 (t_s-t)^{2m}+ a_s^{n-2}\ ,
\label{B8}
\ee
where $m<0$.
Hence, the Casimir energy diverges at the singularity, while
$E_\mathrm{BH}\propto a_s^{n-2}$ takes a finite value
for the singularity time $t_s$, so the bound is clearly violated long
before the singularity.
Then, in order to maintain the validity of the bound (\ref{B1}), one
might assume, as in the Big Rip case,
that GR is not valid near or at the bound. Even if  quantum
effects are included, as was pointed in Ref.\cite{ConformalAnomaly1},
 for this type of singularity the energy density diverges more
rapidly than in the classical case, so that the bound is also
violated in the presence of  quantum effects.

\subsection{Type IV singularity}

For this singularity, the Hubble rate behaves as
\be
\label{III_1}
H = H_1(t) + \left(t_s - t\right)^\alpha H_2(t)\ .
\ee
Here $H_1(t)$ and $H_2(t)$ are regular function and do not vanish at
$t=t_s$.
The constant $\alpha$ is not integer and larger than $1$.
Then the scale factor behaves as
\be
\label{III_2}
\ln a(t) \sim \int dt H_1(t) + \int dt \left(t_s - t\right)^\alpha
H_2(t)\ .
\ee
Near $t=t_s$, the first term dominates and every quantities like
$\rho$, $p$, and $a$ etc. are finite and
therefore the bound (\ref{B1}) would not be violated near the
singularity.

\subsection{Big Bang singularity}

When the matter with $w\geq 0$ coupled with gravity and dominates,
the scale factor behaves as
\be
\label{BB1}
a \sim t^{\frac{2}{n(1+w)}}\ .
\ee
Then there appears a singularity at $t=0$, which may be a Big Bang
singularity.
Although the Big Bang singularity is not a future singularity, we
may consider the bound (\ref{B1}) when $t\sim 0$.
Since $n(1+w)>2$, the energy density behaves as
$\rho \sim a^{-n(1+w)}$ and therefore the Casimir energy behaves as
$E_\mathrm{C} \sim a^{-nw}$. On the other hand, we find
$E_\mathrm{BH} \sim a^{n-2}$.
Then when $n> 2$ or when $n \geq 2$ and $w>0$, $E_\mathrm{C}$
dominates when $a\to 0$, that is, when $t\to 0$,
and the bound (\ref{B1}) is violated. This tells us, as expected, that
 quantum effects become important
in the early universe.

Above, we  have thus explored what happens near the future cosmic
singularities.
We have seen that in general, and with some very special exceptions
on the case of Type II and Type IV,
the bound will be violated if one assumes the validity of GR close to
the singularity. Even if  quantum corrections are assumed, it seems that the bound will  be violated, although in the Big Rip case the singularity may be avoided when quantum effects are incorporated.
It is natural to suggest, in accordance with  Verlinde, that
the bound on the Casimir energy means a finite range for the validity of the classical theory. When this kind of theory becomes saturated,
 some other new quantum gravity effects
have to be taken into account.
We  conclude that the universality of the bound (\ref{B1}) is
not clear and may hold just
for some specific cases, like the radiation dominated Universe.

\section{$F(R)$-gravity and the Cardy-Verlinde formula}

It is known that modified gravity (for general introduction, see
\cite{review}) may be presented in the form of generalized fluid with
inhomogeneous EoS \cite{cap}.
This type of theories which became popular recently may pretend to
 unify  the early-time inflation theory with the theory describing the late-time
acceleration \cite{NO}.
We specify here a modified $F(R)$-gravity modeled  as an effective
fluid and construct the corresponding CV formula for it.
The action that describes $F(R)$-gravity is given by
\be
S=\frac{1}{2\kappa^2}\int d^{n+1}x\sqrt{-g} (F(R) +L_m)\ ,
\label{F1}
\ee
where $L_m$ represents the matter Lagrangian and $\kappa^2=8\pi G$.
The field equations are obtained by varying the action (\ref{F1}) with
respect to the metric $g_{\mu\nu}$,
\be
R_{\mu\nu}F'(R)-\frac{1}{2}g_{\mu\nu}F(R)+g_{\mu\nu}\Box
F'(R)-\nabla_{\mu}\nabla_{\nu}F'(R)
=\kappa^2T^{(m)}_{\mu\nu}\ .
\label{F2}
\ee
Here $T^{(m)}_{\mu\nu}$ is the energy-momentum tensor for the matter
 filling the Universe,
and we have assumed a $1+3$ spacetime for simplicity. For closed
$3+1$ FRW Universe, the modified FRW equations are expressed as
\bea
\frac{1}{2}F(R)-3(H^2+\dot{H})F'(R)+3HF''(R)\dot{R}=\kappa^2
\rho_{m}\ , \nn
-\frac{1}{2}F(R)+\left[3H^2+\dot{H}+\frac{2}{a^2}\right] F'(R)-[(\partial_{tt}F'(R))+2H(\partial_t F'(R))]=\kappa^2 p_m\ ,
\label{F3} \eea
where  primes denote derivatives respect to $R$ and  dots with
respect to $t$. These equations can be rewritten in order to be
comparable with those of standard GR. For such a propose the
geometric terms can be presented as an effective energy-density
$\rho_{F(R)}$ and a pressure $p_{F(R)}$, \bea
H^2+\frac{1}{a^2}=\frac{\kappa^2}{3F'(R)}\rho_m
+\frac{1}{3F'(R)}\left[\frac{RF'(R)-F(R)}{2}-3H\dot{R}F''(R)\right]
\ , \nn 2\dot{H}+3H^2+\frac{1}{a^2}=-\frac{\kappa^2}{F'(R)}p_m
�-\frac{1}{F'(R)}\left[\dot{R}^2F'''(R)+2H\dot{R}F''(R)+\ddot{R}F''(R)
+\frac{1}{2}(F(R)-RF'(R)) \right]\ . \label{F4} \eea Then, an EoS
for the geometric terms can be defined as
$p_{F(R)}=w_{F(R)}\rho_{F(R)}$. We can define an effective
energy-density $\rho=\rho_m/F'(R)+\rho_{F(R)}$ and pressure
$p=p_m/F'(R)+p_{F(R)}$. Hence, for some special cases the formula
for the entropy developed in the second section can be obtained in
$F(R)$-gravity (for an early attempt deriving a  CV formula in a
specific version of $F(R)$-gravity, see \cite{vanzo}). For
example, for an $F(R)$ whose solution gives $\rho\propto
a^{-3(1+w_\mathrm{eff})}$, the formula for the entropy (\ref{10})
is recovered although in general, as in the cases studied above,
no such expression can be given. On the other hand, one could
assume that the geometric terms do not contribute to the matter
sector. Supposing a constant EoS matter fluid, the expression for
the entropy is given by (\ref{10}), although the cosmic Cardy
formula (\ref{20}) has not the same form and in virtue of the
modified first FRW equation (\ref{F3}) the form of the Hubble
entropy $S_\mathrm{H}$, the total energy $E$, and the Bekenstein
energy $E_\mathrm{BH}$, will be very different. It is not easy to
establish correspondence between two such approaches. Note that
using the effective fluid representation the generalized CV
formula may be constructed for any modified gravity.

Now we consider the case where $F(R)$ behaves as
\be
\label{FBB1}
F(R) \sim R^\alpha \ ,
\ee
when the curvature is small or large.
Then if the matter has the EoS parameter $w>-1$, by solving
(\ref{F3}) we find
\be
\label{FBB2}
a \sim \left\{ \begin{array}{cl}
t^{\frac{2\alpha}{n(1+w)}} & \mbox{when}\ \frac{2\alpha}{n(1+w)}> 0
\\
(t_s - t)^{\frac{2\alpha}{n(1+w)}} & \mbox{when}\
\frac{2\alpha}{n(1+w)}< 0
\end{array} \right. .
\ee
Then there may appear a singularity at $t=0$, which corresponds to
the Big Bang singularity,
or at $t=t_s$, which corresponds to the Big Rip singularity.
Since the Casimir energy behaves as $E_\mathrm{C} \sim a^{-nw}$ but
$E_\mathrm{BH} \sim a^{n-2}$,
only when $t\to 0$, $E_\mathrm{C}$ dominates in case that $n> 2$ and
$w\geq 0$ or in case that $n \geq 2$ and $w>0$.
Even in the phantom phase where $\frac{2\alpha}{n(1+w)}< 0$, the
bound (\ref{B1}) is not violated. \\
Let us now consider de Sitter space solution in $F(R)$-gravity
(for review of CV formula in dS or AdS spaces, see \cite{cai}). As
was pointed in Ref.\cite{f(R)deSitter}, almost every function
$F(R)$ admits a de Sitter solution. This can be easily seen from
the first FRW equation in (\ref{F3}). A de Sitter solution is
given by a constant Hubble parameter $H(t)=H_0$; then by inserting
in (\ref{F3}) we obtain the following algebraic equation, \be
3H^2_0=\frac{F(R_0)}{2F'(R_0)}\ . \label{F5} \ee Here
$R_0=12H^2_0$ and the contribution of matter is neglected. Then,
for positive roots  $H_0$ of this equation, the corresponding
$F(R)$ leads to the de Sitter solution which may describe
inflation or dark energy. In this case the formula for the entropy
(\ref{10}) can be reproduced for $w=-1$, and even the universal
bound (\ref{B1}) can hold by taking a critical size of the
Universe. The formula that relates the cosmic bounds in
(\ref{B3a}) is easily obtained also in $F(R)$-gravity for a de
Sitter solution. In such a case one can identify \be
S_\mathrm{H}=\frac{H_0V}{2G}\ , \quad S_\mathrm{B}=\frac{a
V}{24G}\frac{F(R_0)}{F'(R_0)}\ , \quad
S_\mathrm{BH}=\frac{V}{2Ga}\ , \label{F6} \ee which corresponds to
the first FRW equation written as
$S^2_\mathrm{H}+(S_\mathrm{B}-S_\mathrm{BH})^2=S^2_\mathrm{B}$.
Thus, one can conclude that dynamical entropy bounds are not
violated for modified gravity with de Sitter solutions. Note that
quantum gravity effects may be presented also as an effective
fluid contribution. In case when de Sitter space turns out to be
the solution, even with the account of quantum gravity the above
results indicate that dynamical cosmological/entropy bounds are
valid. In other words, the argument indicates the universality of
dynamical bounds. It seems that their violation is caused only by
future singularities if they are not cured by quantum gravity
effects. Note that a large number of modified gravity theories do
not contain  future singularities; they are  cured by higher
derivatives terms.

\section{Discussions}

In summary, we have derived a generalized CV formula for
multicomponent, interacting fluids, generalized in the sense that
an inhomogeneous EoS (including viscous fluid) was assumed. We
also considered  modified $F(R)$-gravity, using its fluid
representation. We showed  that for some special cases the formula
is reduced to the standard CV formula expressing the
correspondence with 2d CFT theory. The dynamical entropy bound for
all above cases was found. The universality of dynamical entropy
bound near  all four types of the future singularity, as well as
the initial Big Bang singularity, was investigated. It was proved
that except from some special cases of Type II and Type IV
singularity the dynamical entropy bound is violated near the
singularity. Taking into  account  quantum effects of conformally
invariant matter does not improve the situation.

One might think that the dynamical entropy bound is universal and that its
violation simply indicates that the situation will be changed with the
introduction of  quantum gravity effects. However,  arguments given below indicate
that it is not the case and that the future singularity is the domain
where all known physical laws and equations are not valid.
Indeed, taking account of quantum effects such as done in section VI does not
improve the situation with respect to non-universality of the dynamical entropy bound.
From another side, it was shown that the dynamical entropy bound is valid
for the de Sitter  solution. Having in mind that quantum gravity
corrections may  always be presented as a generalized effective fluid,
one sees that the dynamical cosmological bound is not valid near the
singularity (even when account is taken of Quantum Gravity). It is only when
modified gravity (with or without
quantum corrections) is regular in the future, like the models of
Refs.\cite{Ralpha,singularity,singularity2} where the future universe is
asymptotically de Sitter, that the dynamical bound
remains valid. Hence, the problem of non-universality of
dynamical entropy bound is related to the more fundamental question about
the real occurrence of a future singularity. It remains a challenge to find
any observational indications for the structure of  the future universe.

\begin{acknowledgments}

This research has been supported in part by MEC (Spain) project
FIS2006-02842 and AGAUR (Catalonia, Spain) 2009SGR-994 (SDO and DSG).
DSG also acknowledgs a grant from MICINN (Spain).
The work by S.N. is supported in part by Global
COE Program of Nagoya University provided by the Japan Society
for the Promotion of Science (G07).

\end{acknowledgments}

\appendix

\section{Account of Casimir effect in the CV formula}

A different way to account for the Casimir energy is to relate the effect
to the single length parameter
in the theory, that is the scale factor, and to assume that we can exploit the
expression for the Casimir effect in a perfectly conducting spherical shell. The method consists essentially in identifying the scale factor with the radius of the shell.
This proposal has been studied in Ref.\cite{Casimir}, and as
shown,  could have some
important effects on the cosmological history.
Here, we want to deduce the expression for the entropy following this
alternative approach.
The expression for the Casimir energy according to \cite{Casimir} can
be expressed as follows
\be
E_\mathrm{C}=\frac{C}{2La}\ ,
\label{C1}
\ee
where $L$ is an auxiliary length that has been introduced due to the
non-dimensional nature of $a$.
This is the same form as the one encountered for a perfectly
conducting shell (see Ref.\cite{CasimirShell}),
where it was found that
\be
C=0.09235\ .
\label{C2}
\ee
Hence, we assume that value of $C$ is much less than the unity, which
 is physically reasonable in view of the conventional feebleness of
the Casimir force.
Also, for simplicity we shall assume a $1+3$ FRW Universe.
The expression (\ref{C2}) corresponds to a Casimir pressure
\begin{equation}
p_c=\frac{-1}{8\pi (La)^2}\frac{\partial E_c}{\partial
(La)}=\frac{C}{8\pi L^4a^4}\ ,
\label{C3}
\end{equation}
and leads consequently to a Casimir energy density $\rho_c \propto
1/a^4$,
which means that the Casimir fluid has an EoS parameter given by
$w_\mathrm{C}=1/3$
in order to obey the energy conservation equation (\ref{4}).
Then, the Casimir energy density and pressure are given by
\begin{equation}
p_c=\frac{C}{8\pi L^4a^4}, \quad \rho_c=\frac{3C}{8\pi L^4a^4}\ .
\label{C4}
\end{equation}
Now we assume that the Universe is filled with a perfect fluid
$p_m=w_m\rho_m$,
where $\rho=E_m/V$.
By following the rescaling properties (\ref{7}), and assuming now
$w_m=1/3$,
the extensive and the Casimir energy are written as a function of the
entropy,
\be
E_m=\frac{\alpha}{4\pi a^w}S^{2/3}\ , \quad
E_\mathrm{C}=\frac{\beta
C}{2\pi L a}S^{2/3}\ .
\label{C5}
\ee
The total energy is $E=E_m+1/2 E_\mathrm{C}$.
Then, the expression for the entropy is very similar to the one
obtained in Ref.\cite{Verlinde},
\be
S=\frac{2\pi L a}{\sqrt{\alpha\beta
C}}\sqrt{E_\mathrm{C}(2E-E_\mathrm{C})}\ ,
\label{C6}
\ee
where $\alpha$ and $\beta$ are arbitrary constants.
We  see that this expression, except from the constants,  is equal
to (\ref{19});
 the constants can be absorbed by $\alpha\beta$.
Then, this approach also supports the formula (\ref{19}), but
like that,
it is not possible to extend the formalism to matter fluids with arbitrary $w_m$.
In that case the expression for the entropy would not be constant
and the first law of thermodynamics would be violated.

Hence, we have obtained a  twofold desription of the Casimir effect.
The Casimir energy appears explicitly in the entropy formula.
Moreover, all quantities are constructed from an effective Casimir
fluid.

\end{document}